\documentclass[doublecol]{epl2}
\usepackage{graphicx,epsfig}
\usepackage{amssymb,amsfonts,amsmath}
\date{\today}

\newcommand{\bk}{{\bf k}}

\newcommand{\rr}{{\bf r}}

\newcommand{\cZ}{{\mathcal Z}}

\title{An ansatz for the exclusion statistics parameters in macroscopic physical systems described by fractional exclusion statistics}
\author{Drago\c s-Victor Anghel}
\institute{Department of Theoretical Physics, National Institute for
  Physics and Nuclear Engineering--''Horia Hulubei'', Str. Atomistilor
  no.407, P.O.BOX MG-6, Bucharest - Magurele, Romania}
\pacs{05.30.-d}{Quantum statistical mechanics}
\pacs{05.30.Ch}{Quantum ensemble theory}
\pacs{05.30.Pr}{Fractional statistics systems (anyons, etc.)}
\abstract{
I introduce an ansatz for the exclusion statistics parameters of fractional exclusion statistics (FES) systems and I apply it to calculate the statistical distribution of particles from both, bosonic and fermionic perspectives. Then, to check the applicability of the ansatz, I calculate the FES parameters in three well-known models: in a Fermi liquid type of system, a one-dimensional quantum systems described in the thermodynamic Bethe ansatz and quasiparticle excitations in the fractional quantum Hall (FQH) systems. The FES parameters of the first two models satisfy the ansatz, whereas those of the third model, although close to the form given by the ansatz, represent an exception. With this ocasion I also show that the general properties of the FES parameters, deduced elsewhere (EPL {\bf 87}, 60009, 2009), are satisfied also by the parameters of the FQH liquid. 
}

\begin{document}
\maketitle

\section{Introduction\label{intro}}

Haldane's concept of fractional exclusion statistics (FES) 
\cite{PhysRevLett.67.937.1991.Haldane} have been recently amended in a 
series of publications 
\cite{JPhysA.40.F1013.2007.Anghel,PhysLettA.372.5745.2008.Anghel,RJP.54.281.2009.Anghel,EPL.87.60009.2009.Anghel}. 
In these publications I showed that in the original formulation of FES 
some basic properties have been overlooked. 
The situation was corrected first by 
introducing a conjecture \cite{JPhysA.40.F1013.2007.Anghel} 
and then by deducing the general, basic properties of the FES parameters 
\cite{EPL.87.60009.2009.Anghel}; it turned out that the conjecture of 
Ref. \cite{JPhysA.40.F1013.2007.Anghel} is just a special case of the 
general conditions deduced in Ref. \cite{EPL.87.60009.2009.Anghel}, which 
allows one to write down an explicitely consistent system 
of equations for the statistical distribution of particles in a FES system. 

Now, that the general properties of the FES parameters are deduced
\cite{EPL.87.60009.2009.Anghel}, the conjecture of Ref. 
\cite{JPhysA.40.F1013.2007.Anghel} looses its status and becomes simply an 
ansatz. This ansatz seems to be quite general and
applies to most of the macroscopic (i.e. quasicontinuous) systems. 
Nevertheless, by the end of this letter I will show an exception. 

In this letter, using the general properties of the direct exclusion statistics 
parameters (see eq. \ref{alphajj} below) I propose an even 
more restrictive (and therefore more conveninet for applications) form of this 
ansatz, by specifying also the form of the direct FES parameters. 
This new ansatz allows me to write the system of equations 
for the statistical distribution of particles in a more clear form and 
to single-out the direct exclusion statistics parameters which 
are most often used in FES calculations. 

In general, the FES equations for the particle distribution are 
used in the bosonic formulation. Here I will write 
and use these equations in both, bosonic and fermionic formulations. 
This allows for direct application of the formalism in systems of 
either bosons or fermions. 

In the end I will give three examples of systems in which FES its manifesting. 
After I calculate their parameters, I show that two of them satisfy the 
ansatz, whereas the third, although quite similar, constitues an exception. 

\section{The general properties of the exclusion statistics parameters} 

Let us assume that we have a system of particles which we divide into 
the species indexed by $i=0,1,\ldots$, each of the species containing 
$N_i$ particles and having $G_i$ ``available single-particle states''. 
Then the number of microscopic configurations for species $i$ is 
\begin{equation}
W_i = \frac{(G_i+N_i)!}{N_i!G_i!}. \label{Wi}
\end{equation}
To introduce FES into the system, we define the \textit{exclusion statistics 
parameters}, $\tilde\alpha_{ij}$, such that at the variation of $N_j$ by 
$\delta N_j$, the number $G_i$ changes by 
$\delta G_i=-\tilde\alpha_{ij}\delta N_j$. The diagonal elements, 
$\tilde\alpha_{ii}$, are called 
\textit{direct exclusion statistics parameters}, 
whereas the nondiagonal ones, $\tilde\alpha_{ij}$, $i\ne j$, are called 
\textit{mutual exclusion statistics parameters}. 

In Ref. \cite{EPL.87.60009.2009.Anghel} I showed that if one of the 
particle species, say species $j$, is divided into 
the subspecies $j_0,j_1,\ldots$, then the new statistics coefficients 
should satisfy the relations 
\begin{subequations}\label{relalphasgen}
\begin{eqnarray}
\tilde\alpha_{ij} &=& \tilde\alpha_{ij_0}=\tilde\alpha_{ij_1}=\ldots,\ 
{\rm for\ any}\ i,\ i\ne j \label{alphaij} \\
\tilde\alpha_{ji} &=& \tilde\alpha_{j_0i}+\tilde\alpha_{j_1i}+\ldots,\ 
{\rm for\ any}\ i,\ i\ne j \label{alphaji} \\
\tilde\alpha_{jj} &=& \tilde\alpha_{j_0j_0}+\tilde\alpha_{j_1j_0}+\ldots 
\nonumber \\
&=& \tilde\alpha_{j_0j_1}+\tilde\alpha_{j_1j_1}+\ldots = \ldots
\label{alphajj}
\end{eqnarray}
\end{subequations}
Here I am using the notations $\tilde\alpha_{ij}$, like in Refs. \cite{JPhysA.40.F1013.2007.Anghel,PhysLettA.372.5745.2008.Anghel,RJP.54.281.2009.Anghel,EPL.87.60009.2009.Anghel}, to make the difference between the 
``extensive'' and ``intensive'' FES parameters that are going to be defined 
below.

The conjecture introduced in Ref. \cite{JPhysA.40.F1013.2007.Anghel} 
stated that in a \textit{macroscopic} physical system described by 
FES there is a division of the system into species, say 
$\{(G_i,N_i)\}_{i=0,1,\ldots}$, so 
that no matter how we divide further these species into subspecies, 
the mutual exclusion statistics parameters are always proportional to 
the dimension of the space on which they act. Concretely, this means that 
for any $i,\ j$, with $i\ne j$, we can write 
$\tilde\alpha_{ij}\equiv a_{ij}G_i$, where 
$a_{ij}$ are constants that depend on the species $i$ and $j$, 
and at any further 
division, say species $i$ is divided into the subspecies 
$i_0,i_1,\ldots$, the new mutual exclusion statistics parameters satisfy 
$\tilde\alpha_{i_kj}=a_{ij}G_{i_k}$ for any $i\ne j$ 
\cite{JPhysA.40.F1013.2007.Anghel}. 
The parameters $\tilde\alpha_{ij}$ are then called the
\textit{extensive} parameters and the parameters $a_{ij}$ are called 
the \textit{intensive} parameters. 
It is easy to check that the extensive parameters satisfy the general 
conditions (\ref{alphaij}) and (\ref{alphaji}). 

In this paper I will extend the ansatz to the direct FES paramteres, 
which should satisfy (\ref{alphajj}).

\section{Ansatz for the FES parameters}

We can make the form of the exclusion statistics parameters that 
satisfy Eqs. (\ref{relalphasgen}) more specific and easier to 
apply to FES calculations if we decompose 
$\tilde\alpha_{ij}$ into a sum of two different types of parameters,
$\tilde\alpha^{(e)}_{ij}$ and $\tilde\alpha^{(s)}_{i}$, by the relation 
\begin{equation}
\tilde\alpha_{ij}=\tilde\alpha^{(e)}_{ij}+\tilde\alpha^{(s)}_{i}\delta_{ij}.
\label{alpha_nd}
\end{equation}
The parameters $\tilde\alpha^{(e)}_{ij}$ are the ``extensive'' 
ones discussed above, only that now, by separating $\tilde\alpha^{(s)}_{i}$, 
we can extend the condition 
\begin{equation}
\tilde\alpha^{(e)}_{ij}\equiv a_{ij}G_i, \label{def_alpha_e}
\end{equation}
also to the case $i=j$. 

The additional parameters, $\tilde\alpha^{(s)}_{i}$, always refer 
to only one species of particles and \textit{are not extensive}. 
If we split the species $i$ into the sub-species $i_0,i_1,\ldots$, then 
by Eqs. (\ref{alphajj}), (\ref{alpha_nd}) and the extensivity property of 
$\tilde\alpha^{(e)}_{ii}$, we obtain 
\begin{equation}
\tilde\alpha_{i_ki_l} = G_{i_k}a_{ii}+\tilde\alpha^{(s)}_{i}\delta_{i_ki_l}
\end{equation}

Typically, in the literature we find exclusion statistics parameters 
of the $(s)$ type (see e.g. 
\cite{PhysRevLett.73.922.1994.Wu,PhysRevLett.73.3331.1994.Murthy,PhysRevLett.74.3912.1995.Sen,PhysRevB.60.6517.1999.Murthy,JPhysB33.3895.2000.Bhaduri,PhysRevLett.86.2930.2001.Hansson,JPA35.7255.2002.Anghel,RomRepPhys59.235.2007.Anghel}).
Therefore in general $\tilde\alpha_{ij}=0$ for any $i\ne j$, so there is 
no mutual statistics in the system. In such a case the thermodynamic 
calculations simplify considerably. Note also that the ideal Fermi gas 
corresponds to $\tilde\alpha^{(s)}_{i}=1$ for any i.

\section{Particle population in the bosonic formulation}

Let us now deduce the equations for the particle population. To avoid 
unphysical (negative or divergent weights) I write the 
number of microscopic configurations as \cite{JPhysA.40.F1013.2007.Anghel}
\begin{eqnarray}
W &=& \prod_{i} \frac{(G_{i}+N_{i}-1+(1-\tilde\alpha^{(s)}_i)\delta N_i
-\sum_{j}\tilde\alpha_{ij}\delta N_{j})!}
{(N_{i}+\delta N_{i})!(G_{i}-1-\alpha^{(s)}_i\delta N_i-\sum_{j}
\tilde\alpha_{ij}\delta N_{j})!} \label{W_ext_pert}\\
&\approx& \prod_{i}\frac{[G_{i}+N_{i}+(1-\tilde\alpha^{(s)}_i)
\delta N_{i}-G_{i}\sum_{j}a_{ij}\delta N_{j}]!}{(N_{i}+\delta N_{i})!
(G_{i}-\tilde\alpha^{(s)}_i\delta N_{i}-G_{i}\sum_{j}a_{ij}\delta N_{j})!} . 
\nonumber 
\end{eqnarray}
which then I plug into the expression for the grandcanonical partition 
function,
\begin{equation}
\cZ = \sum_{\{N_i\}} W(\{N_i\})\exp\left[
\sum_i \beta N_i(\mu_i-\epsilon_i) \right]\,, \label{cZ_gen}
\end{equation}
where $\beta=1/k_B T$, $T$ is the temperature of the system, whereas 
$\mu_i$ and $\epsilon_i$ are the chemical potential and the single-particle 
energy for the particles of species $i$. 
Maximizing $\cZ$ with respect to the populations $n_i\equiv N_i/G_i$, 
I obtain the system of equations 
\begin{equation}
\beta(\mu_i-\epsilon_i)+\ln\frac{[1+n_i]^{1-\tilde\alpha^{(s)}_{i}}}
{n_i} = \sum_{j} G_{j}a_{ji}\ln[1+n_{j}] .
 \label{inteq_for_n1}
\end{equation}
Notice that Eq. (\ref{inteq_for_n1}) is similar to Eq. (18) of Ref. 
\cite{JPhysA.40.F1013.2007.Anghel}, only that by singling out the 
coefficients $\tilde\alpha^{(s)}_{i}$ we could extend the summation on 
the r.h.s. to include also the terms $i=j$. This makes the application of Eq. 
(\ref{inteq_for_n1}) more straightforward than the one in Ref. 
\cite{JPhysA.40.F1013.2007.Anghel}.

If $a_{ij}=0$ for any $i$ and $j$, we recover the typical formulas
for the calculation of particle population without mutual exclusion 
statistics \cite{PhysRevLett.73.922.1994.Wu}, 
\begin{equation}
n=[w(\zeta)+\tilde\alpha^{(s)}_{i}]^{-1},
\end{equation} 
with $w$ and $\zeta$ defined by 
\begin{equation}
w^{\tilde\alpha^{(s)}_{i}}(\zeta)[1+w(\zeta)]^{1-\tilde\alpha^{(s)}_{i}}=
\zeta\equiv \exp[\beta(\epsilon_i-\mu_i)]
\end{equation}

In the quasicontinuous limit, in a phase-space spanned by the single-particle 
states of quantum numbers $\bk$ ($\bk$ is not necessary the wave-number), 
of density of states $\sigma(\bk)$, Eq. (\ref{inteq_for_n1}) transforms into 
\begin{equation}
\beta(\mu_\bk-\epsilon_\bk)+\ln\frac{[1+n_\bk]^{1-\tilde\alpha^{(s)}_{\bk}}}
{n_\bk}=\int \sigma(\bk')\ln[1+n_{\bk'}]a_{\bk'\bk}\,d\bk' .
\label{inteq_for_n2c}
\end{equation}
Equation (\ref{inteq_for_n2c}) is similar to eq. (19) of Ref. 
\cite{JPhysA.40.F1013.2007.Anghel}, if we identify, say 
$\tilde\alpha^{(s)}_{\bk}$ with $\tilde\alpha_{\bk\bk}$ of 
\cite{JPhysA.40.F1013.2007.Anghel}. 
The identification is natural, 
since as the interval $G_i$ becomes smaller, $\tilde\alpha_{ii}$ 
converges to $\tilde\alpha^{(s)}_{ii}$, which stays constant, whereas
$\tilde\alpha^{(e)}_{ii}$ decreases to zero. 

\section{Particle population in the fermionic formulation}

Formula (\ref{Wi}) represents the number of configurations of particles 
of species $i$ in the bosonic formulation of FES 
\cite{PhysRevLett.67.937.1991.Haldane}. This description is not the most 
convenient for example when one describes systems of (interacting) fermions 
in the FES formalism, 
since in such a case $G_i$ represents the difference between the 
number of single-particle states 
and the number of fermions. Therefore in such cases it is easier to work 
directly with the total number of states, $T_i\equiv G_i+N_i-1$, in the 
fermionic description. 
Although the two descriptions are equivalent, let me write down 
the system of equations for the $n_i$s in the fermionic description. 
For this, I write first the 
number of configurations for the species $i$, 
\begin{equation}
W^{(f)}_i = \frac{T_i!}{N_i!(T_i-N_i)!} . \label{Wif}
\end{equation}
We add again small perturbations to the particle numbers and write 
\begin{eqnarray}
W^{(f)} &=& \prod_{i}\left\{\frac{(T_{i}-\tilde\alpha^{(s)}_i\delta N_{i}
-G_{i}\sum_{j}a_{ij}\delta N_{j})!}{[T_{i}-N_{i}-(1+\tilde\alpha^{(s)}_i)
\delta N_{i}-T_{i}\sum_{j}a_{ij}\delta N_{j}]!} \right. \nonumber \\
&& \left.\times\frac{1}{(N_{i}+\delta N_{i})!}\right\} .
\label{W_ext_pert_f}
\end{eqnarray}
I plug (\ref{W_ext_pert_f}) into the expression (\ref{cZ_gen}) for $\cZ$ and, 
by maximization, I get the equations for the particle population, 
$f_i\equiv N_i/T_i$, 
\begin{equation}
\beta(\mu_i-\epsilon_i)+\ln\frac{[1-f_i]^{1+\tilde\alpha^{(s)}_{i}}}
{f_i} = -\sum_{j} G_{j}a_{ji}\ln[1-f_{j}] .
 \label{inteq_for_n1f}
\end{equation}

Introducing the density of states $\sigma^{(f)}(\bk)$ I write Eq. 
(\ref{inteq_for_n1f}) in the quasicontinuous limit, 
\begin{eqnarray}
&& \beta(\mu_\bk-\epsilon_\bk)+\ln\frac{[1-f_\bk]^{1+\tilde\alpha^{(s)}_{\bk}}}
{f_\bk} \label{inteq_for_n2f} \\
&& =-\int \sigma^{(f)}(\bk')\ln[1-f_{\bk'}]a_{\bk'\bk}\,d\bk' .
\nonumber
\end{eqnarray}

\section{Applications}Let me now analyse three interacting particle 
system models in which FES is manifesting and compare their 
FES parameters with the ansatz proposed here. 

\subsection{FES in a system with Fermi liquid type of interaction}

I take again the model of Ref. \cite{PhysLettA.372.5745.2008.Anghel}, 
which is a generalization of the Murthy and Shankar model 
\cite{PhysRevLett.73.3331.1994.Murthy}, widely used in FES \cite{PhysRevLett.73.3331.1994.Murthy,PhysRevB.60.6517.1999.Murthy,JPhysB33.3895.2000.Bhaduri,PhysRevLett.74.3912.1995.Sen,NuclPhysB470.291.1996.Hansson,IntJModPhysA12.1895.1997.Isakov,JPA35.7255.2002.Anghel}. In this model the total energy of the system,
\begin{equation}
E = \sum_i\epsilon_i n_i + \frac{1}{2}\sum_{ij}V_{ij}n_in_j , 
\label{Etotgen}
\end{equation}
is splitted into the quasiparticle energies as 
$E\equiv\sum_i\tilde\epsilon_i n_i$, with 
\begin{equation}
\tilde\epsilon_i = \epsilon_i + \sum_{j=0}^{i-1} V_{ij}n_j +\frac{1}{2}V_{ii}n_i.
\label{epstilgen}
\end{equation}
In Eqs. (\ref{Etotgen}) and (\ref{epstilgen}) $\epsilon_i$ ($i=0,1,\ldots$) 
are single-particle energies, $n_i$ is the population of the state $i$, 
and $V_{ij}$ represent the interaction energy between a 
particle on the state $i$ and a particle on the state $j$. 

Going to the quasi-continuous limit, assuming that the single particle 
energy spectrum has the density of states $\sigma(\epsilon)$ and that the 
interaction energy depends only on the energies, we replace the indices $i$ 
and $j$ by the energies $\epsilon$ and $\epsilon'$ to write 
\begin{equation}
\tilde\epsilon = \epsilon + \int_0^\epsilon V(\epsilon,\epsilon') 
\sigma(\epsilon')n(\epsilon')\,d\epsilon'. 
\label{epstilgenint}
\end{equation}
In what follows I shall assume that the function $\tilde\epsilon(\epsilon)$ 
is bijective and therefore I shall use interchangebly, whenever necessary, 
both $\tilde\epsilon(\epsilon)$ and $\epsilon(\tilde\epsilon)$.

The FES is manifested in this case in the quasiparticle energies, 
$\tilde\epsilon$. In order to describe FES and to calculate its 
parameters, we split the $\tilde\epsilon$ axis into small intervals. In 
general we shal denote such an interval by $\delta\tilde\epsilon$ and 
by this notation we shall assume that it contains the quasiparticle energy 
level $\tilde\epsilon$. 

From Ref. \cite{PhysLettA.372.5745.2008.Anghel} we can directly 
identify the direct exclusion statistics parameters of $(s)$ type, 
\begin{equation}
%
\tilde\alpha^{(s)}_{\tilde\epsilon} =  \frac{V(\epsilon,\epsilon)
\sigma(\epsilon)}{1+ \int_0^\epsilon \frac{\partial V(\epsilon,\epsilon')}
{\partial\epsilon}\sigma(\epsilon')n(\epsilon')\,d\epsilon'} ,
\label{alpha_eps_eps}
\end{equation}
where $\epsilon\equiv\epsilon(\tilde\epsilon)$. 

The mutual exclusion statistics parameters are 
\cite{PhysLettA.372.5745.2008.Anghel}
\begin{equation}
\tilde\alpha_{\delta\tilde\epsilon\delta\tilde\epsilon_i} = \delta\tilde\epsilon
\frac{V(\epsilon,\epsilon_i)+ f(\tilde\epsilon,\tilde\epsilon_i)}
{1+\int_{0}^{\epsilon} \frac{\partial V(\epsilon,\epsilon')}
{\partial\epsilon}\sigma(\epsilon')n(\epsilon')\,d\epsilon'} 
\left[\frac{d\sigma(\epsilon)}{d\epsilon}\right]_{\epsilon(\tilde\epsilon)} ,
\label{alpha_eps_epsp}
\end{equation}
where $\epsilon\equiv\epsilon(\tilde\epsilon)$, 
$\epsilon_i\equiv\epsilon(\tilde\epsilon_i)$, 
$\tilde\epsilon\in\delta\tilde\epsilon$, and
$\tilde\epsilon_i\in\delta\tilde\epsilon_i$. 
For (\ref{alpha_eps_epsp}), the quasiparticles are inserted into the 
interval $\delta\tilde\epsilon_i$ while the variation of the 
number of states is observed in the interval $\delta\tilde\epsilon$.
The function $f(\tilde\epsilon,\tilde\epsilon_i)$ is 
\begin{equation}
f(\tilde\epsilon,\tilde\epsilon_i) 
= \int_{\epsilon_i}^{\epsilon}\frac{\partial 
V(\epsilon,\epsilon')}{\partial\epsilon'}\sigma(\epsilon')n(\epsilon')
\left[\frac{\delta \epsilon'}{\delta\rho(\epsilon_i)}\right]_{\{\rho(\tilde\epsilon)\}}
\,d\epsilon' ,
\label{fepsepsp_def}
\end{equation}
where by $\rho(\tilde\epsilon)\equiv\sigma(\tilde\epsilon)n(\tilde\epsilon)$ 
I denoted the particle density along the $\tilde\epsilon$ axis and 
the notation $\left[\delta \epsilon'/\delta\rho(\epsilon_i)\right]_{\{\rho(\tilde\epsilon)\}}$ represents the functional derivative of $\epsilon'$ 
with respect to the particle density at energy $\tilde\epsilon_i$, when 
we keep fix $\tilde\epsilon(\epsilon')$. Since the number of states in 
the interval $\delta\tilde\epsilon$ is 
$\tilde\sigma(\tilde\epsilon)\delta\tilde\epsilon$, the coefficients 
$a_{\tilde\epsilon\tilde\epsilon_i}$ are 
\begin{equation}
a_{\tilde\epsilon\tilde\epsilon_i} = 
\frac{V(\epsilon,\epsilon_i)+ f(\tilde\epsilon,\tilde\epsilon_i)}
{1+\int_{0}^{\epsilon} \frac{\partial V(\epsilon,\epsilon')}
{\partial\epsilon}\sigma(\epsilon')n(\epsilon')\,d\epsilon'} 
\left[\frac{d\log \sigma(\epsilon)}{d\epsilon}\right]_{\epsilon(\tilde\epsilon)}.
\label{a_eps_epsp}
\end{equation}

Now we are left with the calculation of 
$\tilde\alpha^{(e)}_{\tilde\epsilon\tilde\epsilon}$. For this we first note 
that Eqs. (\ref{alpha_eps_epsp}) and (\ref{a_eps_epsp}) are valid 
for any two disjoint intervals, 
so let us divide the interval $\delta\tilde\epsilon$ into the subintervals 
$\delta\tilde\epsilon_i$, $i=0,1,\ldots,M$, of dimensions 
$\tilde\sigma(\tilde\epsilon_i)\delta\tilde\epsilon_i$, where we always 
maintain the convention that $\tilde\epsilon_i$ belongs to the 
interval $\delta\tilde\epsilon_i$. We pick the energy level 
$\tilde\epsilon_k$ from the interval $\delta\tilde\epsilon_k$ and 
apply Eq. (\ref{alphajj}): 
\begin{equation}
\tilde\alpha_{\delta\tilde\epsilon\delta\tilde\epsilon} = 
\tilde\sigma(\tilde\epsilon_0)\delta\tilde\epsilon_0 
a_{\tilde\epsilon_0\tilde\epsilon_k} +\ldots+
\tilde\alpha_{\delta\tilde\epsilon_k\delta\tilde\epsilon_k}+\ldots+
\tilde\sigma(\tilde\epsilon_M)\delta\tilde\epsilon_M 
a_{\tilde\epsilon_M\tilde\epsilon_k}
\label{til_alpha_desc}
\end{equation}
Making the interval $\delta\tilde\epsilon_k$ small enough as compared to 
$\delta\tilde\epsilon$ and assuming that $\delta\tilde\epsilon$ is also 
small, so that we can use the approximations
$a_{\tilde\epsilon_k\tilde\epsilon_l}=a_{\tilde\epsilon\tilde\epsilon}$ and 
$\tilde\sigma(\tilde\epsilon_k)\equiv\tilde\sigma(\tilde\epsilon)$ for any 
$k,l=1,\ldots,M$, we can 
simplify Eq. (\ref{til_alpha_desc}) to write the general expression 
\begin{equation}
\tilde\alpha_{\delta\tilde\epsilon\delta\tilde\epsilon} = 
a_{\tilde\epsilon\tilde\epsilon}\tilde\sigma(\tilde\epsilon)\delta\tilde\epsilon
+ \tilde\alpha^{(s)}_{\tilde\epsilon} \equiv 
a_{\tilde\epsilon\tilde\epsilon}\tilde\sigma(\tilde\epsilon)\delta\tilde\epsilon
+  \delta(\tilde\epsilon-\tilde\epsilon_i)\tilde\alpha^{(s)}_{\tilde\epsilon},
\label{til_alpha_desc_ea}
\end{equation}
which has the form of the ansatz (\ref{alpha_nd}). For Eq. 
(\ref{til_alpha_desc_ea}), note that $f(\tilde\epsilon,\tilde\epsilon)=0$.

In the simplified models of Refs. \cite{PhysRevLett.73.3331.1994.Murthy,PhysRevB.60.6517.1999.Murthy,JPhysB33.3895.2000.Bhaduri,PhysRevLett.74.3912.1995.Sen,NuclPhysB470.291.1996.Hansson,IntJModPhysA12.1895.1997.Isakov,JPA35.7255.2002.Anghel}, only the direct exclusion statistics parameters have been used and 
$a_{\tilde\epsilon\tilde\epsilon_i}$ was identically zero for any 
$\tilde\epsilon$ and $\tilde\epsilon_i$. We observe now, from eq. 
(\ref{alpha_eps_epsp}), that this happens whenever $d\sigma/d\epsilon=0$. 

Having all the exclusion statistics parameters calculated, one can in 
principle apply Eq. (\ref{inteq_for_n2c}) or (\ref{inteq_for_n2f}), depending 
on the type of particles we have in the system, to calculate 
the particle distribution. 


\subsection{FES in 1D quantum gas in the thermodynamic Bethe ansatz}

The 1D gas of quantum particles in the thermodynamic Bethe ansatz (TBA) 
have been analysed before 
(see e.g. \cite{NewDevIntSys.1995.Bernard,PhysRevB.60.6517.1999.Murthy,PhysRevE.76.61112.2007.Potter}) and is recognized in general as being 
a system which can be described by FES. The only reason for which I 
discuss it again here is to show that its FES parameters are indeed of 
the type (\ref{alpha_nd}) and also because in general a confusion is made in the literature 
between the intensive $a_{ij}$ and extensive $\tilde\alpha_{ij}$ parameters 
and this has to be clarified. 

Therefore let's consider the typical gas of $N$ spinless particles, 
bosons or fermions, 
on a ring of circumference $L$. We assume that the system is 
\textit{nondiffractive} \cite{Sutherland:book} and the asymptotic particle 
wavenumber, $k$, is determined by the equation 
\cite{PhysRevB.56.4422.1997.Sutherland}
\begin{equation}
Lk - \sum_{k'}\theta(k-k') = Lk^{(0)} , \label{ph_shift_eq1}
\end{equation}
where $k^{(0)}\equiv 2\pi I(k)/L$ is the free-particle wavenumber, $I(k)$ 
an integer that depends on $k$, and $\theta(k-k')$ is the phase-shift due 
to the interaction. 

To simplify the notations and to be also in accordance with Refs. 
\cite{Sutherland:book,PhysRevB.56.4422.1997.Sutherland} 
we set the units so that $\hbar=m\equiv1$, where $m$ is the mass of the 
particle. In these units the total number of particles, momentum and 
energy of the system are
\begin{equation}
N=\sum_k 1,\ P=\sum_k k,\ {\rm and}\ E=\frac{1}{2}\sum_k k^2 .
\label{NPE_discret}
\end{equation}

For large systems we transform the summations into integrals and 
define the densities of states, $\sigma(k)$ and $\sigma_0(k^{(0)})$, by 
the relations 
\begin{equation}
D(\delta k) \equiv \sigma(k)\delta k\ {\rm and}\ 
D(\delta k^{(0)}) \equiv \sigma_0(k^{(0)})\delta k^{(0)} \label{def_sigma}
\end{equation}
where $D(\delta k)$ and $D(\delta k^{(0)})$ are the numbers of states 
in the small intervals $\delta k$ and $\delta k^{(0)}$, respectively. 
If $\delta k$ and $\delta k^{(0)}$ are related by Eq. (\ref{ph_shift_eq1}), 
then $D(\delta k)=D(\delta k^{(0)})$. Obviously, 
$\sigma_0(k^{(0)})=L/(2\pi)$ (if we impose periodic boundary conditions on 
$k^{(0)}$) \cite{Sutherland:book,PhysRevB.56.4422.1997.Sutherland}.

The populations of the single particle levels, $n_0(k^{(0)})$ and 
$n(k)$, are defined as 
\begin{equation}
N(\delta k) \equiv n(k)\sigma(k)\delta k = 
N(\delta k^{(0)}) \equiv n_0(k^{(0)})\sigma_0(k^{(0)})\delta k^{(0)} \label{def_N}
\end{equation}
where $N(\delta k)=N(\delta k^{(0)})$ is the number of particles in the 
interval $\delta k$ or $\delta k^{(0)}$, with $\delta k$ and
related by Eq. (\ref{ph_shift_eq1}). Moreover, since both, the number 
of particles and the number energy levels, are the same in the $\delta k$ 
and $\delta k^{(0)}$ intervals, we have the identity $n(k)=n_0[k^{(0)}(k)]$. 

In accordance with the notations in the literature 
\cite{NewDevIntSys.1995.Bernard,PhysRevB.56.4422.1997.Sutherland} I 
introduce here also the particle density, 
$\rho(k)\equiv\sigma(k)n(k)/L=N(\delta k)/(L\delta k)$.

In the new notations, Eq. (\ref{ph_shift_eq1}) becomes a self-consistent 
equation for $k$,
\begin{equation}
k = k^{(0)}(k)+\int \theta(k-k')\rho(k')\,dk' ,
\label{Eq_k_cont1}
\end{equation}
from which we can calculate \cite{PhysRevB.56.4422.1997.Sutherland}
\begin{eqnarray}
\frac{dk^{(0)}}{dk} &\equiv& 1-\int\theta'(k-k')
\rho(k')\,dk' , \label{dk0dk} 
\end{eqnarray}
where $\theta'(k)\equiv d\theta(k)/dk$. Plugging Eq. 
(\ref{dk0dk}) into (\ref{def_sigma}), we get the density of states 
%
\begin{eqnarray}
\sigma(k) &=& \frac{L}{2\pi}\left\{1-
\int\theta'(k-k')\rho(k')\,dk'\right\} . \label{defsigmak1} 
\end{eqnarray}
%

The FES is manifesting in the system because of the dependence of $\sigma$ on 
$\rho$: the variation of $\rho(k_i)$ (changing the number of particles at the 
level $k_i$) produces, in principle, a change of the density of states 
$\sigma(k)$, at any $k$. 
To determine the coefficients of the exclusion statistics we calculate 
the variation of $\sigma(k)$ at the 
variation of $\rho(k_i)$, i.e. we calculate the functional derivative
\begin{equation}
\frac{\delta\sigma(k)}{L\delta\rho(k_i)} = -\frac{1}{2\pi}\theta'(k-k_i) 
\label{dsigmadrho}
\end{equation}
Therefore if we split now the $k$ axis into small intervals, with  
$\delta k$ the interval around $k$ and $\delta k_i$ the interval around 
$k_i$, then the variation of $N(\delta k_i)$ by $\delta N(\delta k_i)$ 
produces a variation of $D(\delta k)$ equal to 
\begin{equation}
\delta D(\delta k) = \frac{\delta\sigma(k)}{L\delta\rho(k_i)}\delta k = 
-\frac{1}{2\pi}\theta'(k-k_i)\delta k\delta N(\delta k_i)
\label{deltaD1}
\end{equation}
From the FES formula, 
$\delta D(\delta k) = -\tilde\alpha_{\delta k\delta k_i}\delta N(\delta k_i)$,
we obtain imediately 
\begin{subequations}\label{mutuals}
\begin{equation}
\tilde\alpha_{\delta k\delta k_i} = \frac{1}{2\pi}\theta'(k-k_i)\delta k 
\label{tildealpha}
\end{equation}
and therefore the exclusion statistics parameter 
$\tilde\alpha_{\delta k\delta k_i}$ \textit{is proportional to the dimension 
of the space on which it acts}, $D(\delta k)$. 

Comparing Eq. (\ref{tildealpha}) with Eq. (\ref{def_alpha_e}) we 
get 
\begin{equation}
a_{kk_i} = \frac{\theta'(k-k_i)}{2\pi\sigma(k)} . \label{ckki} 
\end{equation}
\end{subequations}

From Eqs. (\ref{mutuals}) all the TBA thermodynamics 
follows in general, by direct application 
of the formalism presented before, in either 
bosonic or fermionic formulations, as we shall see imediately. 

If the particles are \textit{bosons}, we plug $a_{kk_i}$ into 
(\ref{inteq_for_n2c}), setting $\tilde\alpha^{(s)}_{k}$, we obtain after 
some simplifications 
\begin{equation}
\epsilon(k) = \epsilon^{(0)}_k - \mu -\frac{k_BT}{2\pi} 
\int\log[1-e^{-\beta\epsilon(k')}]\theta'(k'-k)\,dk' . \label{eq_eps_B}
\end{equation}
In Eq. (\ref{eq_eps_B}) I used the notation $\epsilon^{(0)}_k\equiv k^2/2$ 
and I defined the quasiparticle energy, 
$\epsilon(k)$, by 
$n_k \equiv \{\exp[\beta\epsilon(k)]-1\}^{-1}$.

If the particles are fermions, we define the quasiparticle energy by 
$\exp[-\beta\epsilon(k)]=n_k/(1-n_k)$ and we plug Eq. (\ref{ckki}) into 
Eq. (\ref{inteq_for_n2f}). In this way we recover imediately 
the TBA equation, 
\begin{equation}
\epsilon(k)  = \epsilon^{(0)}(k)-\mu + \frac{k_BT}{2\pi}\int
\log[1+e^{-\beta\epsilon(k')}] \theta'(k'-k)\,dk' , \label{TBAB}
\end{equation}
similarly to Eq. (\ref{eq_eps_B}). 


For the delta function interaction potential between the particles, 
$V(x)=2c\delta(x)$ ($x$ being the distance between the particles), 
the phase shift is $\theta(k)=-2\arctan(k/c)$, which gives 
\begin{equation}
a_{kk_i} = -\frac{1}{\pi\sigma(k)}\cdot\frac{c}{c^2+(k-k_i)^2}
\end{equation}
and therefore 
\begin{equation}
\tilde\alpha^{(e)}_{\delta k\delta k_i} = -\frac{\delta k}{\pi}\cdot
\frac{c}{c^2+(k-k_i)^2},
\end{equation}
while 
\begin{equation}
\tilde\alpha^{(s)}_{\delta k}=0
\end{equation}
for any $k$. 

If the particle-particle interaction is $V(x)=\lambda(\lambda-1)/x^2$, 
then the phase shift is $\theta(k)=\pi(\lambda-1){\rm sgn}(k)$, where 
${\rm sgn}$ is the sign function. From this we obtain 
\begin{equation}
a_{kk_i} = \frac{(\lambda-1)}{\sigma(k)}\delta(k-k_i) . 
\label{akkiCS}
\end{equation}
Note that in this case we do not have any ``extensive'' mutual exclusion 
statistics parameters and therefore we may write 
$\tilde\alpha_{\delta k\delta k_i}\equiv\tilde\alpha^{(s)}_{k}
\delta_{\delta k\delta k_i}=(\lambda-1)\delta_{\delta k\delta k_i}$. 

In both cases we recover the ansatz (\ref{alpha_nd}). 

\subsection{Fractional Quantum Hall Effect}

Another system which is traditionally related to the FES is the 
fractional quantum Hall effect (FQHE) \cite{PhysRevLett.67.937.1991.Haldane,PhysRevLett.73.922.1994.Wu,PhysRevLett.77.3423.1996.Su,cond-mat9706030.Wu}. 

In a Laughlin $1/m$-liquid, with $m$ an odd integer, at any finite 
temperature there are quasiparticle vortex-like excitations, $N_+$ and 
$N_-$, corresponding 
to quasi electrons and quasi holes. The numbers of quasi excitations are 
related to the number of flux quanta in the system, $N_\phi=eBA/hc$ 
(where $B$ is the magnetic flux and $A$ is the area of the sample), and 
electron number, $N_e$, by the relation 
\cite{PhysRevLett.77.3423.1996.Su,cond-mat9706030.Wu}, 
\begin{equation}
N_\phi = mN_e+N_--N_+
\end{equation}
For single particle ocupancy (only one quasi-excitation in the system), the 
number of available states for each of these types of excitations 
is $G_-=G_+=N_e$, while for general $N_+$ and $N_-$ we have 
\begin{subequations}\label{FQHE1}
\begin{eqnarray}
G_+ &=& \frac{1}{m}N_\phi-\alpha_{++}N_+-\alpha_{+-}N_- 
\label{G+FQHE} \\
G_- &=& \frac{1}{m}N_\phi-\alpha_{-+}N_+-\alpha_{--}N_- 
\label{G-FQHE}
\end{eqnarray}
\end{subequations}
where $\alpha_{ij}$ ($i,j=+,-$) are the FES parameters. 
Although maybe there is still no consensus regarding the values of 
$\alpha_{ij}$, which differ for the different liquid models used 
to describe the FQHE (see e.g. \cite{cond-mat9706030.Wu}), for 
concreteness I shall adopt here the bosonic vortex scheme 
\cite{PhysRevLett.77.3423.1996.Su,cond-mat9706030.Wu,PhysRevB.55.6727.1997.Isakov}, with 
\begin{subequations}\label{BVSFQHE}
\begin{eqnarray}
&& \alpha_{++} = 2-1/m,\ \alpha_{+-}=1/m, \label{BVSFQHE1} \\
&& \alpha_{-+} = -1/m,\ \alpha_{--}=1/m, \label{BVSFQHE2}
\end{eqnarray}
\end{subequations}
alhough this is not important for our discussion. 

The point is that although the fractional quantum Hall liquid (HQHL) is a 
macroscopic system, apparently the FES parameters 
of this system do not obey the general relations (\ref{relalphasgen}): 
there are only two species of quasiparticles--the quasi-electrons and the 
quasi-holes, with degenerate energy levels--and the $\alpha$'s are fixed 
by (\ref{BVSFQHE}), so we cannot (aparently) split 
the species into further subspecies. Still, relations (\ref{relalphasgen}) 
are deduced on very general grounds, so they should be valid. 

We have a puzzle. 

The solution of this puzzle is straightforward and may be 
obtained only by macroscopic considerations. 
For this we observe that FQHL being a macroscopic system, one 
can always 
divide its area, $A$, into smaller areas, $A_i$, $i=0,1,\ldots$, and in 
each of the smaller subsystems the FQHL has the same properties, but 
with scalled 
number of electrons, $N_{e_i}$, flux quanta, $N_{\phi_i}$, quasi-electrons, 
$N_{+_i}$, quasi-holes, $N_{-_i}$, and $G_{\pm_i}=G_\pm\cdot A_i/A$. 
In equilibrium, 
$N_{e_i}=N_{e}\cdot A_i/A$, $N_{\phi_i}=N_{\phi}\cdot A_i/A$, 
$N_{+_i}=N_{+}\cdot A_i/A$ and $N_{-_i}=N_{-}\cdot A_i/A$. 
In each of these subsystems, the same relations, (\ref{FQHE1}) and 
(\ref{BVSFQHE}), are valid for the quantities $N_{e_i}$, $N_{\phi_i}$, 
$N_{+_i}$, $N_{-_i}$, and $G_{\pm_i}=G_\pm\cdot A_i/A$. Notice with this 
occasion that there is no ``mutual'' statistics 
between different sub-systems, $i$ and $j$ ($i\ne j$). 
Therefore, by continuing to split $A$ into samller and smaller 
areas, we may eventually end-up with a coarse-graied surface of elemetary 
areas, $\delta A(x,y)$, where the $x$ and $y$ are the two-dimensional 
coordinates on the surface of the FQHL, and with the FES parameters, 
$\alpha_{ij}[\delta A(x,y)]$ ($i,j=+,-$) that act always on 
\textit{the same elementary area}, $(\delta A(x,y))$. 

The puzzle is only aparent. Since the system is macroscopic, 
therefore extensive, the quantities $\alpha_{ij}$ are local and
rigorously should be written as 
$\alpha_{ij}(\rr,\rr')\equiv \alpha_{ij}\delta(\rr-\rr')$, for any 
$i,j=+,-$, where by $\rr$ and $\rr'$ I denoted two position vectors 
$(x,y)$ and $(x',y')$ in the plane. 
This form of FES parameters obey the genreal relations (\ref{relalphasgen}), 
as they should, but constitue an exception to the ansatz (\ref{alpha_nd}) 
proposed in the beggining, due to the fact that the off-diagonal elements 
$\alpha_{+-}$ and $\alpha_{-+}$ are not extensive, but are also proportional 
to $\delta(\rr-\rr')$. 

In the reverse process, if we ``glue'' together all the elementary areas, 
$\delta A(\rr)$, we reobtain the original system, of area $A$, two 
species of particles, $N_+$ and $N_-$, and the overall FES paramters 
(\ref{BVSFQHE}). 

\section{Conclusions}

In this paper I introduced an ansatz for the FES parameters and with it 
I calculated the statistical distribution of particles from both, bosonic and 
fermionic perspectives. This ansatz allowed me to write the system of equations 
for the statistical distribution of particles in a more clear form and 
to single-out the direct exclusion statistics parameters which 
are most often used in FES calculations

Then I took three examples: a Fermi liquid type of system, a one-dimensional 
integrable quantum system and a fractional quantum Hall (FQH) system. I 
calculated the FES parameters for these systems and I showed that those 
of the first two systems obey the ansatz proposed here, whereas those 
of the third one do not. With this 
ocasion I also showed that if one takes properly in to account the 
extensivity of a FQH system, then its FES paramters also obey 
the general conditions deduced in Ref. \cite{EPL.87.60009.2009.Anghel}.

\section{Acknowledgements}

I thank Dr. Ovidiu P\^a\c tu and Prof. J. M. Leinaas for motivating 
discussions and the staff members of the Bogoliubov Laboratory of 
Theoretical Physics of JINR-Dubna, Russia, for hospitality and 
interesting discussions during the visit in which I did most part of this 
work. The work was partially supported by the NATO grant, 
EAP.RIG 982080. 


\end{document}